# Tuple spaces implementations and their efficiency

Vitaly Buravlev, Rocco De Nicola, Claudio Antares Mezzina

*IMT School for Advanced Studies Lucca, Piazza S. Francesco, 19, 55100 Lucca, Italy*

SUMMARY

Among the paradigms for parallel and distributed computing, the one popularized with Linda, and based on tuple spaces, is one of the least used, despite the fact of being intuitive, easy to understand and to use. A tuple space is a repository, where processes can add, withdraw or read tuples by means of atomic operations. Tuples may contain different values, and processes can inspect their content via pattern matching. The lack of a reference implementation for this paradigm has prevented its widespread. In this paper, first we perform an extensive analysis of a number of actual implementations of the tuple space paradigm and summarise their main features. Then, we select four such implementations and compare their performances on four different case studies that aim at stressing different aspects of computing such as communication, data manipulation, and cpu usage. After reasoning on strengths and weaknesses of the four implementations, we conclude with some recommendations for future work towards building an effective implementation of the tuple space paradigm.

## 1. INTRODUCTION

Distributed computing is getting increasingly pervasive, with demands from various application domains and with highly diverse underlying architectures that range from the multitude of tiny devices to the very large cloud-based systems. Several paradigms for programming parallel and distributed computing have been proposed so far. Among them we can list: distributed shared memory [28] (with shared objects and tuple spaces [20] built on it) remote procedure call (RPC [7]), remote method invocation (RMI [30]) and message passing [1] (with actors [4] and MPI [5] based on it). Nowadays, the most used paradigm seems to be message passing while the least popular one seems to be the one based on tuple spaces that was proposed by David Gelernter for the Linda coordination model [19].

As the name suggests, message passing permits coordination by allowing exchanges of messages among distributed processes, with message delivery often mediated via brokers. In its simplest incarnation, message-passing provides a rather low-level programming abstraction for building distributed systems. Linda instead provides a higher level of abstraction by defining operations for synchronization and exchange of values between different programs that can share information by accessing common repositories named *tuple spaces*. The Linda interaction model provides time and space decoupling [18], since tuple producers and consumers do not need to know each other.

The key ingredient of Linda is a small number of basic operations which can be embedded into different programming languages to enrich them with communication and synchronization facilities. Three atomic operations are used for writing (`out`), withdrawing (`in`), and reading (`rd`) tuples into/from a tuple space. Another operation `eval` is used to spawn new processes. The operations for reading and withdrawing select tuples via *pattern-matching* and block if the wanted data are not available. Writing is instead performed by asynchronous output of the information for interacting partners. Figure 1 illustrates an example of a tuples space with different, structured, values. For example tuple $\langle \text{"goofy"}, 4, 10.4 \rangle$ is produced by a process via the $\texttt{out}(\langle \text{"goofy"}, 4, 10.4 \rangle)$ operation, and is read by the operation $\texttt{rd}(\langle \text{"goofy"}, \_, \_ \rangle)$ after pattern-matching: that is the process reads any tuple of three elements whose first one is exactly the string "goofy". Moreover, tuple $\langle 10, \langle \ldots \rangle \rangle$ is consumed (atomically retracted) by operation $\texttt{in}(\langle 10, x \rangle)$ which consumes a tuple whose first element is 10 and binds its second element (whatever it is) to the variable $x$. Patterns are sometimes called *templates*.

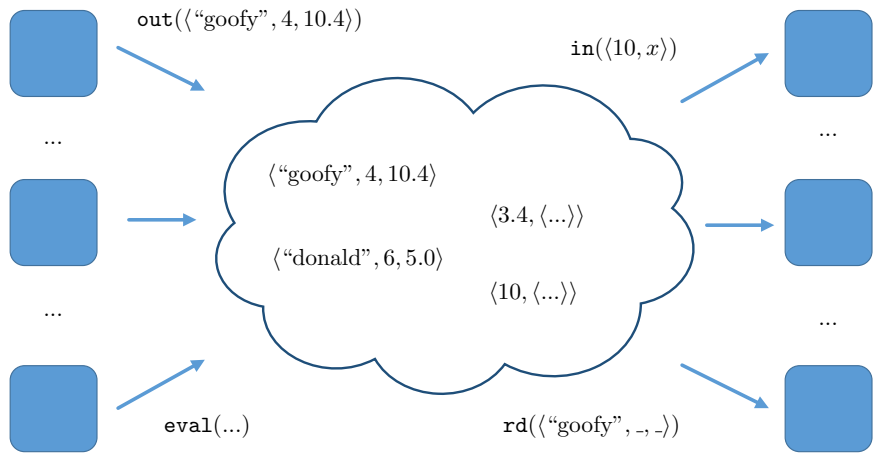

Figure 1. A tuple space

The simplicity of this coordination model makes it very intuitive and easy to use. Some synchronization primitives, e.g. semaphores or barrier synchronization, can be implemented easily in Linda (cf. [10], Chapter 3). Unfortunately, Linda's implementations of tuple spaces have turned out to be quite inefficient, and this has led researchers to opt for different approaches such OpenMP or MPI, which are nowadays offered, as libraries, for many programming languages. When considering distributed applications, the limited use of the Linda coordination model is also due to the need of guaranteeing consistency of different tuple spaces. In fact, in this case, control mechanisms that can significantly affect scalability are needed [12].

In our view, tuple spaces can be effectively exploited as a basis for the broad range of the distributed applications with different domains (from lightweight applications to large cloud based systems). However, in order to be effective, we need to take into account that performances of a tuple space system may vary depending on the system architecture and on the type of interaction between its components. Although the concept of tuple spaces is rather simple, the main challenge to face when implementing it is to devise the best data structure to deal with a possibly distributed *multiset* of tuples, where operations on it (e.g. pattern-matching, insertion and removal) are optimized. Moreover, it has to support efficient parallel tuples' processing and data distribution. Depending on how these aspects are implemented, performances of an application can be positively or negatively affected.

The aim of this paper is to examine the current implementations of tuple spaces and to evaluate their strengths and weaknesses. We plan to use this information as directions for the building more efficient implementation of distributed tuple space.

We start by cataloging the existing implementations according to their features, then we focus on the most recent Linda based systems that are still maintained while paying specific attention to those offering decentralized tuples space. We compare the performances of the selected systems on four different case studies that aim at stressing different aspects of computing such as communication, data manipulation, and cpu usage. After reasoning on strength and weakness of the four implementations, we conclude with some recommendations for future work towards building an effective implementation of the tuple space paradigm.

The rest of the paper is organized as follows. In Section 2 we survey existing tuple spaces systems and choose some of them for the practical examination. The description of case studies, main principles of their implementation, and the results of the experiments are presented in Section 3. Section 4 concludes the paper by collecting some remarks and highlighting some directions for future work. This paper is a revised and extended version of [8]; it contains an additional case study, the thorough evaluation of a new tuple space system and more extensive experiments.

## 2. TUPLE SPACE SYSTEMS

Since the first publication on Linda [20], there have been a plenty of implementations of its coordination model in different languages. Our purpose is to review the most significant and recent ones, that are possibly still maintained, avoiding toy implementations or the one shot paper implementations. To this end, we have chosen: JavaSpaces [26] and TSpaces [24] which are two industrial proposals of tuple spaces for Java; GigaSpaces [21] which is a commercial implementation of tuple spaces; Tupleware [3] featuring an adaptive search mechanism based on communication history; Grinda [9], Blossom [33], DTuples [22] featuring distributed tuple spaces; LuaTS [23] which mixes reactive models with tuple spaces; Klaim [15] and MozartSpaces [14] which are two academic implementations with a good record of research papers based on them.

In this Section, first we review the above mentioned tuple space systems by briefly describing each of them, and single out the main features of their implementations, then we summarise these features in Table I. Later, we focus on the implementations that enjoy the characteristics we consider important for a tuple space implementation: code mobility, distribution of tuples and flexible tuples manipulation. All tuple space systems are enumerated in order they were first mentioned in publications.

**Blossom.** Blossom [33] is a C++ implementation of Linda which was developed to achieve high performance and correctness of the programs using the Linda model. In Blossom all tuple spaces are homogeneous with a predefined structure that demands less time for type comparison during the tuple lookup. Blossom was designed as a distributed tuple space and can be considered as a distributed hash table. To improve scalability each tuple can be assigned to a particular place (a machine or a processor) on the basis of its values. The selection of the correspondence of the tuple and the machine is based on the following condition: for every tuple the field access pattern is defined, that determines which fields always contain value (also for templates); values of these fields can be hashed to obtain a number which determines the place where a tuple has to be stored. Conversely, using the data from the template, it is possible to find the exact place where a required tuple is potentially stored. Prefetching allows a process to send an asynchronous (i.e. non-blocking) request for a tuple and to continue its work while the search is performed. When the requested tuple is needed, if found, it is received without waiting.

**TSpaces.** TSpaces [24] is an implementation of the Linda model developed at the IBM Almaden Research Center. It combines asynchronous messaging with database features. TSpaces provides a transactional support and a mechanism of tuple aging. Moreover, the embedded mechanism for access control to tuple spaces is based on access permission. It checks whether a client is able to perform specific operations in the specific tuples space. Pattern matching is performed using either standard `equals` method or `compareTo` method. It can use also SQL-like query that allows matching tuples regardless of their structure, e.g., ignoring the order in which fields are stored.

**Klaim.** KLAIM [15] (A Kernel Language for Agents Interaction and Mobility) is an extension of Linda supporting distribution and processes mobility. Processes, like any other data, can be moved from one locality to another and can be executed at any locality. Klava [6] is a Java implementation of KLAIM that supports multiple tuple spaces and permits operating with explicit localities where processes and tuples are allocated. In this way, several tuples can be grouped and stored in one locality. Moreover, all the operations on tuple spaces are parameterized with a locality. The emphasis is put also on access control which is important for mobile applications. For this reason, KLAIM introduces a type system which allows checking whether a process is allowed to perform specific operations at specific localities.

**JavaSpaces.** JAVASPACES [26] is one of the first implementations of tuple spaces developed by Sun Microsystems. It is based on a number of Java technologies (e.g., Jini and RMI). Like TSPACES, JAVASPACES supports transactions and a mechanism of tuple aging. A tuple, called entry in JAVASPACES, is an instance of a Java class and its fields are the public properties of the class. This means that tuples are restricted to contain only objects and not primitive values. The tuple space is implemented by using a simple Java collection. Pattern matching is performed on the byte-level, and the byte-level comparison of data supports object-oriented polymorphism.

**GigaSpaces.** GIGASPACES [21] is a contemporary commercial implementation of tuple spaces. Nowadays, the core of this system is GIGASPACES XAP, a scale-out application server; user applications should interact with the server to create and use their own tuple space. The main areas where GIGASPACES is applied are those concerned with big data analytics. Its main features are linear scalability, optimization of RAM usage, synchronization with databases and several database-like features such as complex queries, transactions, and replication.

**LuaTS.** LUATS [23] is a reactive event-driven tuple space system written in Lua. Its main features are the associative mechanism of tuple retrieving, fully asynchronous operations and the support of code mobility. LUATS provides centralized management of the tuple space which can be logically partitioned into several parts using indexing. LUATS combines the Linda model with the event-driven programming paradigm. This paradigm was chosen to simplify program development since it allows avoiding the use of synchronization mechanisms for tuple retrieval and makes more transparent programming and debugging of multi-thread programs. Tuples can contain any data which can be serialized in Lua. To obtain a more flexible and intelligent search of tuples, processes can send to the server code that once executed returns the matched tuples. The reactive tuple space is implemented as a hash table, in which data are stored along with the information supporting the reactive nature of that tuple space (templates, client addresses, callbacks and so on).

**MozartSpaces.** MOZARTSPACES [14] is a Java implementation of the space-based approach [27]. The implementation was initially based on the eXtensible Virtual Shared Memory (XVSM) technology, developed at the Space Based Computing Group, Institute of Computer Languages, Vienna University of Technology. The basic idea of XVSM is related to the concept of *coordinator*: an object defining how tuples (called entries) are stored. For the retrieval, each coordinator is associated with a *selector*, an object that defines how entries can be fetched. There are several predefined coordinators such as FIFO, LIFO, Label (each tuple is identified by a label, which can be used to retrieve it), Linda (corresponding to the classic tuple matching mechanism), Query (search can be performed via a query-like language) and many others. Along with them, a programmer can define a new coordinator or use a combination of different coordinators (e.g. FIFO and Label). MozartSpaces provides also transactional support and a role based access control model [13].

**DTuples.** DTuples [22] is designed for peer-to-peer networks and based on distributed hash tables (DHT), a scalable and efficient approach. Key features of DHT are autonomy and decentralization. There is no central server and each node of the DHT is in charge of storing a part of the hash table and of keeping routing information about other nodes. As the basis of the DTH's implementation DTuples uses FreePastry*. DTuples supports transactions and guarantees fault-tolerance via replication mechanisms. Moreover, it supports multi tuple spaces and allows for two kinds of tuple space: *public* and *subject*. A public tuple space is shared among all the processes and all of them can perform any operation on it. A subject tuple space is a private space accessible only by the processes that are bound to it. Any subject space can be bound to several processes and can be removed if no process is bound to it. Due to the two types of tuple spaces, pattern matching is specific for each of them. Templates in the subject tuple space can match tuples in the same subject tuple space and in the common tuple space. However, the templates in the common tuple space cannot match the tuple in the subject tuple spaces.

**Grinda.** Grinda [9] is a distributed tuple space which was designed for large scale infrastructures. It combines the Linda coordination model with grid architectures aiming at improving the performance of distributed tuple spaces, especially with a lot of tuples. To boost the search of tuples, Grinda utilizes spatial indexing schemes (X-Tree, Pyramid) which are usually used in spatial databases and Geographical Information Systems. Distribution of tuple spaces is based on the grid architecture and implemented using structured P2P network (based on Content Addressable Network and tree based).

**Tupleware.** Tupleware [3] is specially designed for array-based applications in which an array is decomposed into several parts each of which can be processed in parallel. It aims at developing a scalable distributed tuple space with good performances on a computing cluster and provides simple programming facilities to deal with both distributed and centralized tuple space. The tuple space is implemented as a hashtable, containing pairs consisting of a key and a vector of tuples. Since synchronization lock on Java hashtable is done at the level of the hash element, it is possible to access concurrently to several elements of the table. To speed up the search in the distributed tuple space, the system uses an algorithm based on the history of communication. Its main aim is to minimize the number of communications for tuples retrieval. The algorithm uses *success factor*, a real number between 0 and 1, expressing the likelihood of the fact that a node can find a tuple in the tuple space of other nodes. Each instance of Tupleware calculates success factor on the basis of previous attempts and first searches tuples in nodes with greater success factor.

In order to compare the implementations of the different variants of Linda that we have considered so far, we have singled out two groups of criteria.

The first group refers to criteria which we consider fundamental for any tuple space system:

`eval` **operation** This criterion denotes whether the tuple space system has implemented the `eval` operation and, therefore, allows using code mobility. It is worth mentioning that the original `eval` operation was about asynchronous evaluation and not code mobility, but in the scope of a distributed tuple space, it makes programming data manipulation more flexible.

---

*FreePastry is an open-source implementation of Pastry, a substrate for peer-to-peer applications (http://www.freepastry.org/FreePastry/).

|  | **JSP** | **TSP** | **GSP** | **TW** | **GR** | **BL** | **DTP** | **LTS** | **KL** | **MS** |
|---|---|---|---|---|---|---|---|---|---|---|
| `eval` operation | | | | | | | | ✓ | ✓ | |
| Tuple clustering | | | ? | ✓ | | | | ✓ | | |
| No domain specificity | ✓ | ✓ | ✓ | | ✓ | ✓ | ✓ | | ✓ | ✓ |
| Security | | ✓ | ✓ | | | | | | ✓ | |
| Distributed tuple space | | | ? | ✓ | ✓ | ✓ | ✓ | | ✓ | ✓ |
| Decentralized management | | | | ✓ | ✓ | | ✓ | ✓ | ✓ | ✓ |
| Scalability | | | ✓ | ✓ | ✓ | | ✓ | | | |

JavaSpaces (**JSP**), TSpaces (**TSP**), GigaSpaces (**GSP**), Tupleware (**TW**), Grinda (**GR**), Blossom (**BL**), DTuples (**DTP**), LuaTS (**LTS**), Klaim (**KL**) MozartSpaces (**MS**)

Table I. Results of the comparison

**Tuples clustering** This criterion determines whether some tuples are grouped by particular parameters that can be used to determine where to store them in the network.

**Absence of domain specificity** Many implementations have been developed having a particular application domain in mind. On the one hand, this implies that domain-specific implementations outperform the general purpose one, but on the other hand, this can be considered as a limitation if one aims at generality.

**Security** This criterion specifies whether an implementation has security features or not. For instance, a tuple space can require an authorization and regulate the access to its tuples, for some of them, the access can be limited to performing specific operations (e.g. only writes or read).

The second group, of criteria, gathers features which are desirable for any fully distributed implementation that runs over a computer network, does not rely on a single node of control or management and is scalable.

**Distributed tuple space** This criterion denotes whether tuple spaces are stored in one single node of the distributed network or they are spread across the network.

**Decentralized management** Distributed systems rely on a node that controls the others or the control is shared among several nodes. Usually, systems with the centralized control have bottlenecks which limit their performance.

**Scalability** This criterion implies that system based on particular Linda implementation can cope with the increasing amount of data and nodes while maintaining acceptable performance.

Table I summarises the result of our comparison: ✓ means that the implementation enjoys the property and ? means that we were not able to provide an answer due to the lack of source code and/or documentation.

After considering the results in Table I, to perform our detailed experiments we have chosen: TUPLEWARE which enjoys most of the wished features; KLAIM since it offers distribution and code mobility; MOZARTSPACES since it satisfies two important criteria of the second group (fully distribution) and is one of the most recent implementation. Finally, we have chosen GIGASPACES because it is the most modern among the commercial systems; it will be used as a yardstick to compare the performance of the others. We would like to add that DTUPLES has not been considered for the more detailed comparison because we have not been able to obtain its libraries or source code and that GRINDA has been dropped because it seems to be the less maintained one.

In all our implementations of the case studies, we have structured the systems by assigning each process a local tuple space. Because GIGASPACES is a centralized tuple space, in order to satisfy this rule we do not use it as centralized one, but as distributed: each process is assigned its own tuple space in the GIGASPACES server.

## 3. EXPERIMENTS

In order to compare four different tuple space systems we consider four different case studies: *Password search*, *Sorting*, *Ocean model* and *Matrix multiplication*. We describe them below.

*3.1. Introducing case studies*

The first case study is of interest since it deals with a large number of tuples and requires to perform a huge number of write and read operations. This helps us understand how efficiently an implementation performs operations on local tuple spaces with a large number of tuples. The second case study is computation intensive since each node spends more time for sorting elements than on communicating with the others. This case study has been considered because it needs structured tuples that contain both basic values (with primitive type) and complex data structures that impacts on the speed of the inter-process communication. The third case has been taken into account since it introduces particular dependencies among nodes, which if exploited can improve the application performances. This was considered to check whether adapting a tuple space system to the specific inter-process interaction pattern of a specific class of the applications could lead to significative performance improvements. The last case study is a communication-intensive task and it requires much reading on local and remote tuple spaces. All case studies are implemented using the master-worker paradigm [10] because among other design patterns (e.g., Pipeline, SPMD, Fork-join) [25] it fits well with all our case studies and allows us to implement them in a similar way. We briefly describe all the case studies in the rest of this subsection.

**Password search.** The main aim of this application is to find a password using its hashed value in a predefined "database" distributed among processes. Such a database is a set of files containing pairs (password, hashed value). The application creates a master process and several worker processes (Figure 2): the master keeps asking the workers for passwords corresponding to a specific hashed values, by issuing tuples of the form:

$$\langle \text{``search\_task''}, dd157c03313e452ae4a7a5b72407b3a9 \rangle$$

Each worker first loads its portion of the distributed database and then obtains from the master a task to look for the password corresponding to a hash value. Once it has found the password, it sends the result back to the master, with a tuple of the form:

$$\langle \text{``found\_password''}, dd157c03313e452ae4a7a5b72407b3a9, 7723567 \rangle$$

For multiple tuple space implementations, it is necessary to start searching in one local tuple space and then to check the tuple spaces of other workers. The application terminates its execution when all the tasks have been processed and the master has received all required results.

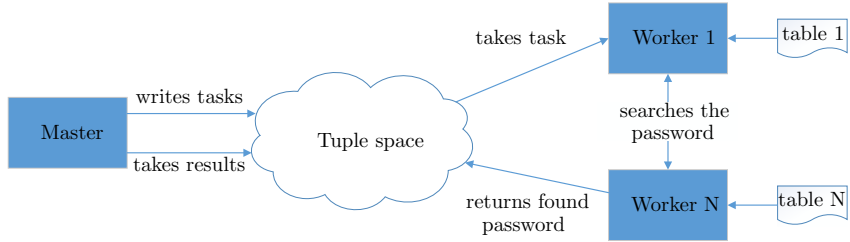

Figure 2. Schema of the case study *Password seach*

**Sorting.** This distributed application consists of sorting an array of integers. The master is responsible for loading initial data and for collecting the final sorted data, while workers are directly responsible for the sorting. At the beginning, the master loads predefined initial data to be sorted and sends them to one worker to start the sorting process. Afterward, the master waits for the sorted arrays from the workers: when any sub-array is sorted the master receives it and when all sub-arrays are collected builds the whole sorted sequence. An example of the sorting is shown in Figure 3 where we have the initial array of 8 elements. For the sake of simplicity, the Figure illustrates the case in which arrays are always divided into equal parts and sorted when the size of each part equals 2 elements, while in the real application it is parametric to a threshold. In the end, we need to reconstruct a sorted array from already sorted parts of smaller size.

The behavior of workers is different; when they are instantiated, each of them starts searching for the unsorted data in local and remote tuple spaces. When a worker finds a tuple with unsorted data, it checks whether the size of such data is below the predetermined threshold and in such a case it computes and sends the result to the master; then it continues by searching for other unsorted data. Otherwise, the worker splits the array into two parts: one part is stored into its local tuple space while the other is processed.

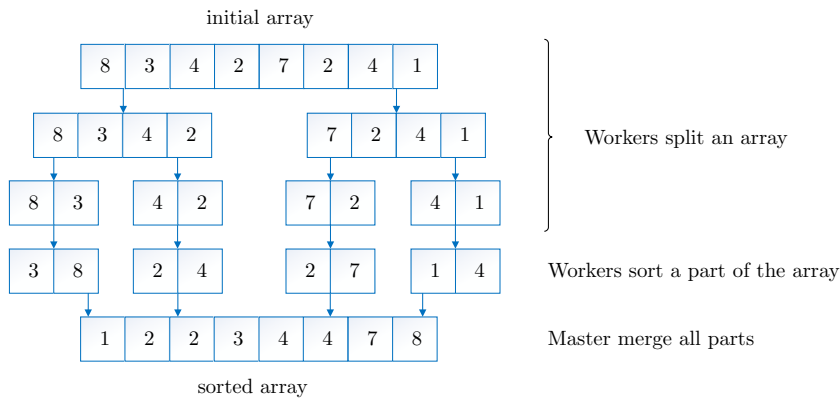

Figure 3. Schema of the case study *Sorting*

**Ocean model.** The ocean model is a simulation of the enclosed body of water that was considered in [3]. The two-dimensional (2-D) surface of the water in the model is represented as a 2-D grid and each cell of the grid represents one point of the water. The parameters of the model are current velocity and surface elevation which are based on a given wind velocity and bathymetry. In order to parallelize the computation, the whole grid is divided into vertical panels (Figure 4), and each worker owns one panel and computes its parameters. The parts of the panels, which are located on the border between them are colored. Since the surface of the water is continuous, the state of each point depends on the states of the points close to it. Thus, the information about bordering parts of panels should be taken into account. The aim of the case study is to simulate the body of water during several time-steps. At each time-step, a worker recomputes the state (parameters) of its panel by exploiting parameters of the adjacent panels.

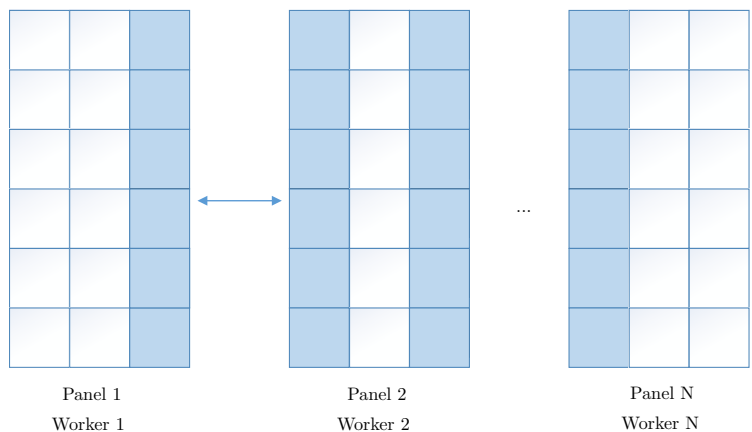

Figure 4. Schema of the case study *Ocean model*

The missions of the master and workers are similar to the previous case studies. In the application the master instantiates the whole grid, divides it into parts and sends them to the workers. When all the iterations are completed, it collects all parts of the grid. Each worker receives its share of the grid and at each iteration it communicates with workers which have adjacent grid parts in order to update and recompute the parameters of its model. When all the iterations are completed, each worker sends its data to the master.

**Matrix multiplication.** The case study is designed to multiply two square matrices of the same order. The algorithm of multiplication [31] operates with rows of two matrices A and B and put the result in matrix C. The latter is obtained via subtasks where each row is computed in parallel. At the $j$-th step of a task the $i$-th task, the element, $a_{ij}$, of A is multiplied by all the elements of the $j$-th row of B; the obtained vector is added to the current $i$-th row of C. The computation stops when all subtasks terminate. Figure 5 shows how the first row of C is computed if A and B are $2 \times 2$ matrices. In the first step, the element $a_{1,1}$ is multiplied first by $b_{1,1}$ then by $b_{1,2}$, to obtain the first partial value of the first row. In the second step, the same operation is performed with $a_{1,2}$, $b_{2,1}$ and $b_{2,2}$ and the obtained vector is added to the first row of C thus obtaining its final value.

Initially, the master distributes the matrices A and B among the workers. In our case study we have considered two alternatives: (i) the rows of both A and B are spread uniformly, (ii) the rows of A are spread uniformly while B is entirely assigned to a single worker. This helped us in understanding how the behavior of the tuple space and its performances change when only the location of some tuples changes.

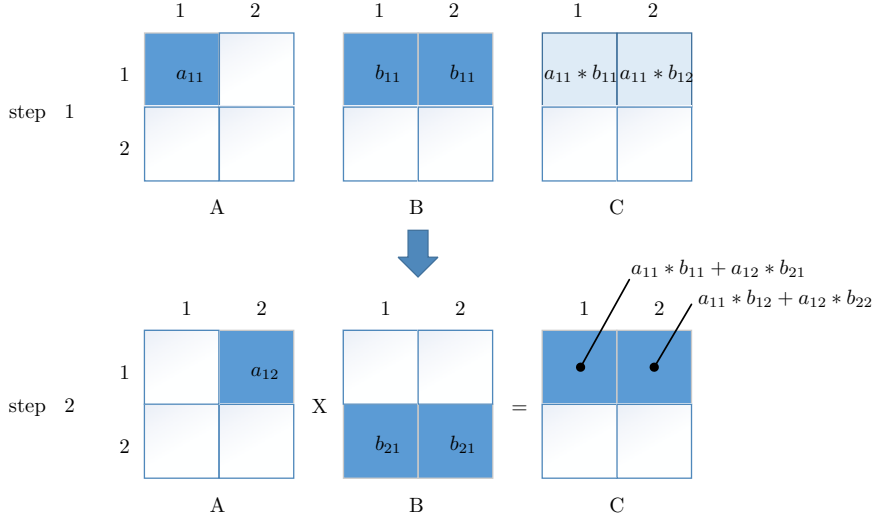

Figure 5. Schema of the case study *Matrix mutltiplication*

*3.2. Implementing the case studies*

All the chosen implementations are Java-based, the most used language (according to TIOBE Programming Community Index [32] and [11]). It guarantees also the possibility of performing better comparisons of the time performances exhibited by the different tuple systems which could significantly depend on the chosen language.

Another key-point of using the same programming language for all the implementations, is that case studies can be written as *skeletons*: the code remains the same for all the implementations while only the invocations of different library methods do change. In order to implement the skeleton for the case study, we have developed several wrappers for each of four tuple space systems we have chosen and each wrapper implements interface `ITupleSpace`. This interface defines the basic operations of a tuple space (e.g. initialization/destruction, I/O operations and so on). Since the tuple space systems have a different set of operations on tuple space we have chosen those operations which have the same semantics for all systems and can be unified. It is worth to notice that all I/O operations on a tuple space are wrapped and placed in the class *TupleOperation*.

To show the principle of *skeleton* we take a look at the skeleton of the case study *Password search*. Master behavior is implemented by class `DistributedSearchMaster` shown in Listing 1. The complete master code along with the workers one can be found in Appendix A. Listing 1 contains just an excerpt of the code, reporting just the salient parts of this case study. The class (Listing 1) is generic with respect to an object extending class `ITupleSpace`. This class is a wrapper/interface for the different tuple space systems.

The logic of the master and worker process follows the description of the case study given above. The master process first initializes its local tuple space of the system given by parameter of the class (lines 21-22). After that, it waits until all the workers are ready (line 25), load all data they need and starts logging the execution time (line 26-29). Then the process creates the tasks for the workers and waits for the results (lines 37-41). When all the results are gathered, the master notifies the workers that they can finish their work, stops counting the time of execution and saves the data of profiling (lines 44-53). Let us note, that thanks to the use of generics (e.g lines 32-34), the master code abstracts away from how the different tuple spaces systems implement the operations on tuples.

There is a difference on how the tuple spaces systems implement the search among distributed tuple spaces. TUPLEWARE has a built-in operation with notification mechanism:

Listing 1: Password search. Excerpt of the master process

```java
public class DistributedSearchMaster<T extends ITupleSpace> {

  // class fields
  private Object masterTSAddress;
  private int numberOfWorkers;
  private int numberOfElements;
  private Class tupleSpaceClass;

  // tuple type
  public static Object[] searchTupleTemplate =
    new Object[] {String.class, String.class, String.class};

  public DistributedSearchMaster(Object masterTSAddress, int numberOfElements,
    int numberOfWorkers, Class tupleSpaceClass) {
    //initialising fields
  }

  public void passwordSearchMaster() {

      // initialize a local tuple space
      T masterTS = getInstanceOfT(tupleSpaceClass);
      masterTS.startTupleSpace(masterTSAddress, numberOfWorkers, true);

      // wait when all workers will be available
      waitForAllWorkers(masterTS);
      TupleLogger.begin("Master::TotalRuntime");

      // wait when all workers will load tables with data
      waitForDataLoad(masterTS);

      // spread the current test key
      TupleOperations.writeTuple(masterTS, masterTS,
          masterTS.formTuple("SearchTuple",
        new Object[]{"search", "master_key", DProfiler.testKey},
        searchTupleTemplate), true, false);

      // create tasks and write it into local tuple space of the master
      int numberOfTasks = 100;
      taskCreation(masterTS, numberOfTasks);

      // wait and get accomplished tasks
      ArrayList<Object[]> foundTuples = getAccomplishedTasks(masterTS,
          numberOfTasks);

      // send to worker − "finish its work"
      for(int i =0; i< numberOfWorkers; i++) {
      TupleOperations.writeTuple(masterTS, masterTS,
        masterTS.formTuple("SearchTuple", new Object[]{"search_task", "",
            completeStatus},
        searchTupleTemplate), true, false);
      }
      TupleLogger.end("Master::TotalRuntime");

      // write data of the test
      DProfiler.writeTestKeyToFile(DProfiler.testKey);
      TupleLogger.writeAllToFile(DProfiler.testKey);

    // wait until all workers will end
      Thread.sleep(10000);
      masterTS.stopTupleSpace();
    }
}
```

it searches in local and remote tuple spaces at once (i.e. in broadcast fashion) and then waits for the notification that the desired tuple has been found. Mimicking the behavior of this operation for the other tuple space systems requires to continuously check each tuple space until the required tuple is found.

*3.3. Assessment Methodology*

All the conducted experiments are parametric with respect to two values. The first one is the number of workers $w$, $w \in \{1, 5, 10, 15\}$. This parameter is used to test the scalability of the different implementations. The second parameter is application specific, but it aims at testing the implementations when the workload increases.

- *Password search* we vary the number of the entries in the database (10000, 1000000, 1 million passwords) where it is necessary to search a password. This parameter directly affects the number of local entries each worker has. Moreover, for this case study the number of passwords to search was fixed to 100.
- *Sorting* case, we vary the size of the array to be sorted (100000, 1 million, 10 million elements). In this case the number of elements does not correspond to the number of tuples because parts of the array are transferred also as arrays of smaller size.
- *Ocean model* we vary the grid size (300, 600 and 1200) which is related to the computational size of the initial task.
- *Matrix multiplication* we vary the order of a square matrix (50, 100).

*Remark 1* (*Execution environment*). Our tests were conducted on a server with 4 processors Intel Xeon E5620 (4 cores, 12M Cache, 2.40GHz, Hyper-Threading Technology ) with 32 threads in total, 40 GB RAM, running Ubuntu 14.04.3. All case studies are implemented in Java 8 (1.8.0).

**Measured metrics.** For the measurement of metrics we have created a profiler which is similar to Clarkware Profiler[†]. Clarkware Profiler calculates just the average time for the time series, while ours also calculates other statistics (e.g., standard deviation). Moreover, our profiler was designed also for analyzing tests carried out on more than one machine. For that reason, each process writes raw profiling data on a specific file; all files are then collected and used by specific software to calculate required metrics.

We use the manual method of profiling and insert methods `begin(label)` and `end(label)` into program code surrounding parts of the code we are interested in order to begin and stop counting time respectively. For each metrics the label is different and it is possible to use several of them simultaneously. This sequence of the actions can be repeated many times and eventually, all the data are stored on disk for the further analysis.

Each set of experiments has been conducted 10 times with randomly generated input and computed an average value and a standard deviation of each metrics. To extensively compare the different implementations, we have collected the following measures:

**Local writing time:** time required to write one tuple into a local tuple space.

**Local reading time:** time required to read or take one tuple from a local tuple space using a template. This metrics checks how fast pattern matching works.

**Remote writing time:** time required to communicate with a remote tuple space and to perform write operation on it.

**Remote reading time:** time required to communicate with a remote tuple space and to perform read or take operation on it.

**Search time:** time required to search a tuple in a set of remote tuple spaces.

---

[†]The profiler was written by Mike Clark; the source code is available on GitHub: https://github.com/akatkinson/Tupleware/tree/master/src/com/clarkware/profiler

**Total time:** total execution time. The time does not include an initialization of tuple spaces.

**Number of visited nodes:** number of visited tuple spaces before a necessary tuple was found.

*3.4. Experimental Results*

Please notice that all plots used in the paper report results of our experiments in a logarithmic scale. When describing the outcome, we have only used those plots which are more relevant to evidence the difference between the four tuple space systems.

**Password search.** In Figures 6-7 is reported the trend of the total execution time as the number of workers and size of considered database increase. In Figure 6 the size of the database is 100 thousand entries, while Figure 7 reports the case in which the database contains 1 million of elements. From the plot, it is evident that GigaSpaces exhibits better performances than the other systems.

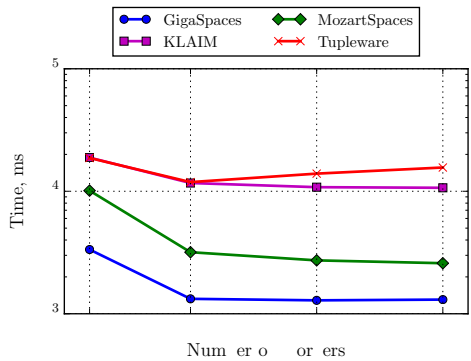

Figure 6. Password search. Total time (100 thousand passwords)

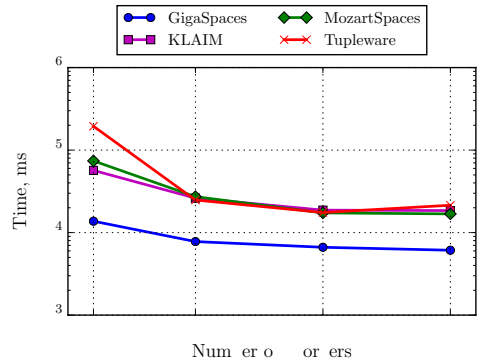

Figure 7. Password search. Total time (1 million passwords)

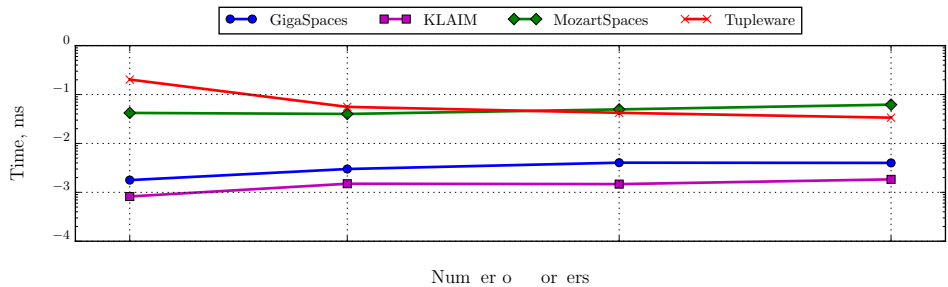

Figure 8. Password search. Local writing time (1 million passwords)

Figure 8 depicts the local writing time for each implementation with different numbers of workers. As we can see, by increasing the number of workers (that implies reducing the amount of local data to consider), the local writing time decreases. This is more evident for Tupleware, which really suffers when a big number of tuples (e.g. 1 million) is stored in a single local tuple space. The writing time of Klaim is the lowest among other systems and does not change significantly during any variation in the experiments. The local writing time of MozartSpaces remains almost the same when the number of workers increases. Nonetheless, its local time is bigger with respect to the other systems, especially when the number of workers is equal or greater than 10.

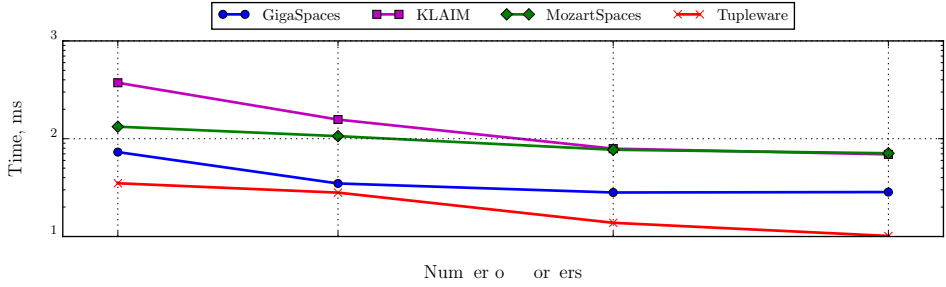

Figure 9. Password search. Local reading time (1 million passwords)

Local reading time is shown in Figure 9 and KLAIM is the one that exhibits the worst performance for searching in a local space. Indeed, if there is just one worker, the local reading time is 10 times greater than TUPLEWARE. This can be ascribed to the pattern matching mechanism of KLAIM which is less effective than others. By increasing the number of workers the difference becomes less evident and approximately equal to MOZARTSPACES time that does not change considerably but always remains much greater than the time of TUPLEWARE and GIGASPACES. Since this case study requires little synchronization among workers, performance improves when the level of parallelism (the number of workers) increases.

To better clarify the behaviors of KLAIM and TUPLEWARE (the only implementations for which the code is available) we can look at how local tuple spaces are implemented. KLAIM is based on `Vector` that provides a very fast insertion with the complexity $O(1)$ if it is performed at the end of the vector and a slow lookup with the complexity $O(n)$. TUPLEWARE has `Hashtable` as a container for tuples but the use of it depends on specific types of templates that we do not satisfy in our skeleton-implementation (namely, for TUPLEWARE first several fields of the template should contain values that is not always the case for our skeleton). Therefore, in our case all the tuples with passwords are stored in one vector meaning that the behavior is similar to one of KLAIM. However, TUPLEWARE insert every new tuple in the begging of the vector that slows down the writing and a simplified comparison (based on a comparison of strings) for the lookup that makes it faster.

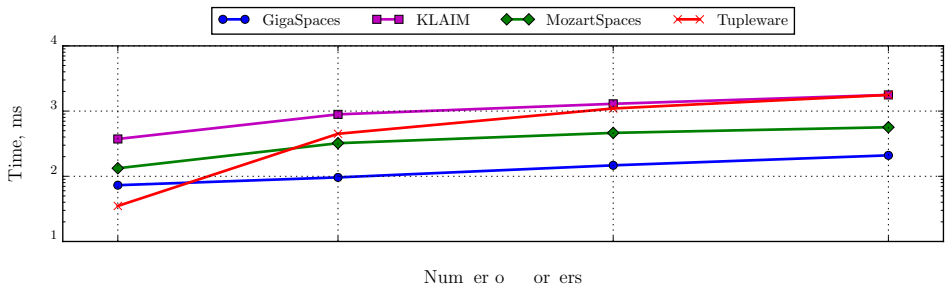

Figure 10. Password search. Search time (1 million passwords)

The search time is similar to the local reading time but takes into account searching in remote tuple spaces. When considering just one worker, the search time is the same as the reading time in a local tuple space, however, when the number of workers increases the search time of TUPLEWARE and KLAIM grows faster than the time of GIGASPACES. Figure 10 shows that GIGASPACES and MOZARTSPACES are more sensitive to the number of tuples than to the number of accesses to the tuple space.

Summing up, we can remark that the local tuple spaces of the four systems exhibit different performances depending on the operation on them: the writing time of KLAIM is always significantly smaller than the others, while the pattern matching mechanism of Tupleware allows for faster local searching. The performance of MozartSpaces mostly depends on the number of involved workers: it exhibits average time for local operations when one worker is involved, while it shows the worst time with 15 workers.

**Sorting.** Figure 11 shows that GigaSpaces exhibits significantly better execution time when the number of elements to sort is 1 million. As shown in Figure12 when 10 million elements are considered and several workers are involved, Tupleware exhibits a more efficient parallelization and thus requires less time. For the experiment with more than 10 workers and array size of 10 million, we could not get results of MozartSpaces because some data were lost during the execution making it not possible to obtain the sorted array. This is why in Figures 12, 13 and 14 some data of MozartSpaces are missing. This is caused by a race condition bug when two processes try to simultaneously write data to a third process since we experienced this data loss when sorted sub-array were returned to the master. Other tuple space systems have not shown such a misbehavior.

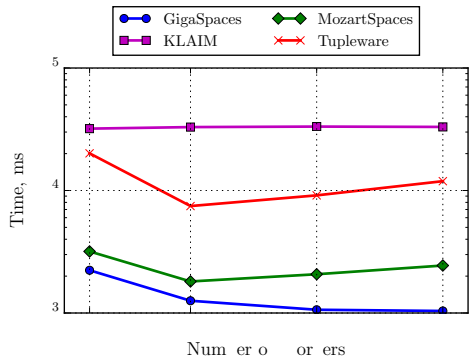
Figure 11. Sorting. Total time (1 million elements)

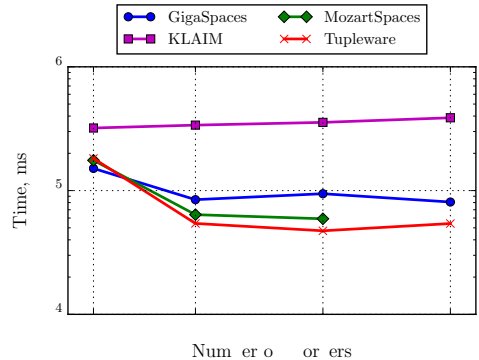
Figure 12. Sorting. Total time (10 million elements)

This case study is computation intensive but requires also an exchange of structured data and, although in the experiments a considerable part of the time is spent for sorting, we noticed that performance does not significantly improve when the number of workers increases.

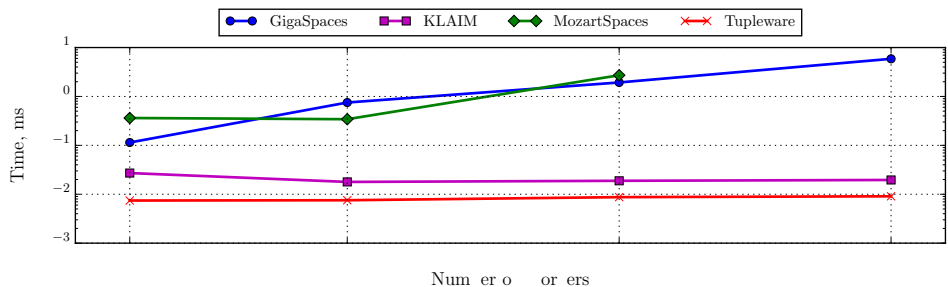

Figure 13. Sorting. Local writing time (10 million elements)

The performance of Klaim is visibly worse than others even for one worker. In this case, the profiling of the Klaim application showed that a considerable amount of time was spent for passing initial data from the master to the worker. Inefficient implementation of data transmission seems to be the reason the total time of Klaim differs from the total time of Tupleware.

By comparing Figures 8 and 13, we see that when the number of workers increases, GigaSpaces and Klaim suffer more from synchronization in the current case study than in the previous one.

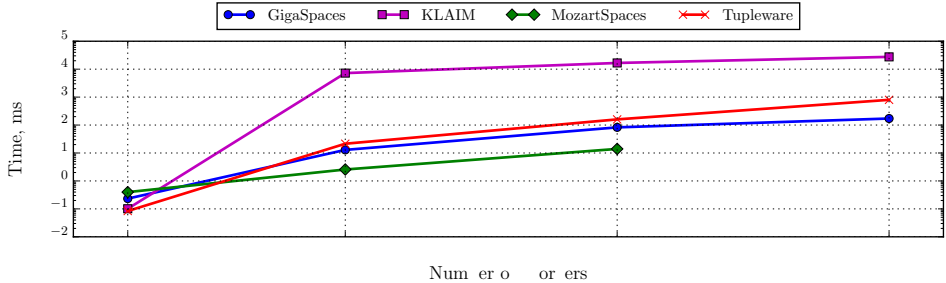

Figure 14. Sorting. Search time (10 million elements)

As shown in Figure 14, search time directly depends on the number of the workers and grows with it. Taking into account that Klaim and Tupleware spend more time accessing remote tuple space, GigaSpaces suffers more because of synchronization. Klaim has the same problem, but its inefficiency is hampered by data transmission cost.

**Ocean model.** This case study was chosen to examine the behavior of tuple space systems when specific patterns of interactions come into play. Out of the four considered systems, only Tupleware has a method for reducing the number of visited nodes during search operation which helps in lowering search time. Figure 15 depicts the number of visited nodes for different grid sizes and a different number of workers (for this case study in all figures we consider only 5, 10, 15 workers because for one worker generally tuple space is not used). The curve depends weakly on the size of the grid for all systems and much more on the number of workers. Indeed, from Figure 15 we can appreciate that Tupleware performs a smaller number of nodes visits and that when the number of workers increases the difference is even more evident[‡].

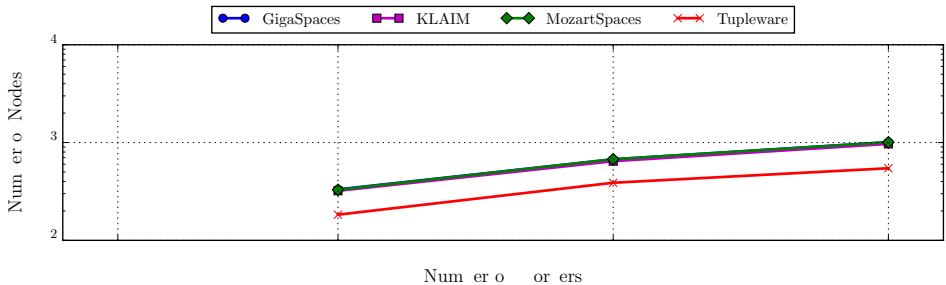

Figure 15. Ocean model. Number of visited nodes (grid size is 1200)

The difference in the number of visited nodes does not affect significantly the total time of execution for different values of the grid size (Figure 16-17) mostly because the case study requires many read operations from remote tuple spaces (Figure 18).

As shown in Figure 18 the time of remote operation varies for different tuple space systems. For this case study, we can neglect the time of the pattern matching and consider that this time is equal to the time of communication. For Klaim and Tupleware these times were similar and significantly greater than those of GigaSpaces and MozartSpaces. Klaim and

---

[‡]Figure 15, the curves for Klaim and GigaSpaces are overlapping and purple wins over blue.

Tupleware communications rely on TCP and to handle any remote tuple space one needs to use exact addresses and ports. GigaSpaces, that has a centralized implementation, most likely does not use TCP for data exchange but relies on a more efficient memory-based approach. The communication time of MozartSpaces is in the middle (in the plot with logarithmic scale) but close to GigaSpaces by its value: for GigaSpaces this time varies in the range of 0.0188 to 0.0597 ms, for MozartSpaces in the range of 2.0341 to 3.0108 ms, and for Tupleware and Klaim it exceeds 190 ms. Therefore, as it was mentioned before, GigaSpaces and MozartSpaces implements a read operation differently from Tupleware and Klaim and it is more effective when working on a single host.

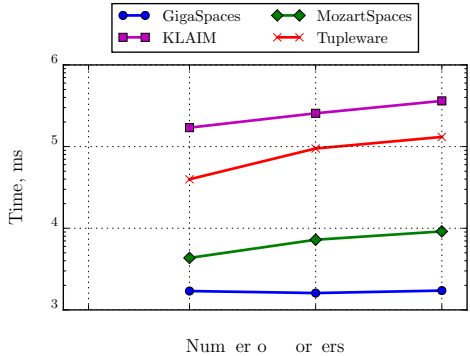

Figure 16. Ocean model. Total time (grid size is 600)

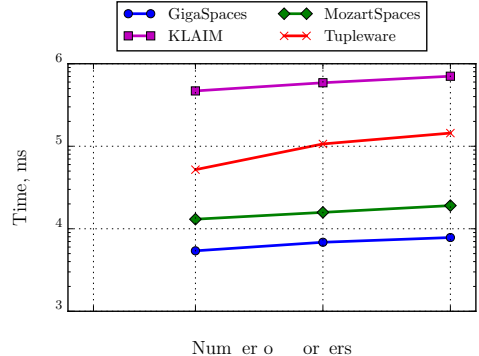

Figure 17. Ocean model. Total time (grid size is 1200)

Figure 16 provides evidence of the effectiveness of Tupleware when its total execution time is compared with the Klaim one. Indeed, Klaim visits more nodes and spends more time for each read operation, and the difference increases when the grid size grows and more data have to be transmitted as it is shown in Figure 17.

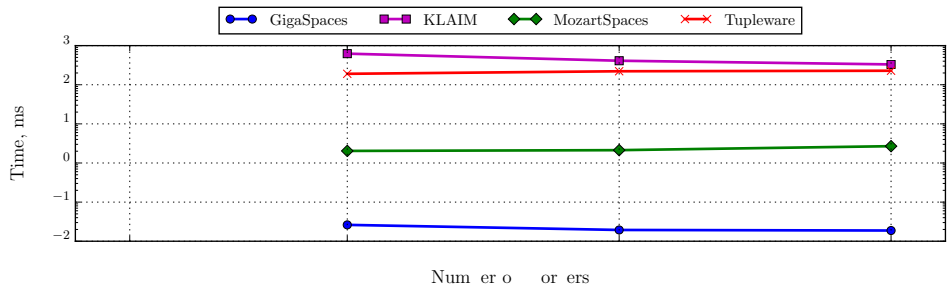

Figure 18. Ocean model. Remote reading time (grid size is 1200)

**Matrix multiplication.** This case study mostly consists of searching tuples in remote tuple spaces, and this implies that the number of remote read operations is by far bigger than the other operations. Therefore, GigaSpaces and MozartSpaces outperform other tuple space systems total execution time (Figure 19).

As discussed in Section 3.1 we will consider two variants of this case study: one in which matrix B is uniformly distributed among the workers (as the matrix A), and one in which the whole matrix is assigned to one worker. In the following plots, solid lines correspond to the experiments with uniform distribution and dashed lines correspond to ones with the second type of distribution (we name series with this kind of distribution with a label ending with B-1).

Figure 20 depicts the average number of the nodes that it is necessary to visit in order to find a tuple for each worker. When considering experiments with more than one worker all tuple space systems except TUPLEWARE demonstrate similar behavior: the total time almost coincides for both types of the distribution. However, for the uniform distribution TUPLEWARE exhibits always greater values and for the second type of distribution, the values are significantly lower. The second case reaffirms the results of the previous case study because in this case all workers know where to search the rows of the matrix B almost from the very beginning that leads to the reduction of the amount of communication, affects directly the search time (Figure 21) and, in addition, implicitly leads to the lower remote reading time (Figure 22, the remote reading time is not displayed for one worker because only the local tuple space of the worker is used). In contrast, for the uniform distribution TUPLEWARE performs worse because of the same mechanism which helps it in the previous case: when it needs to iterate over all the rows one by one it always starts the checking from the tuple spaces which were already checked at the previous time and which do not store required rows. Therefore, every time it checks roughly all tuple spaces.

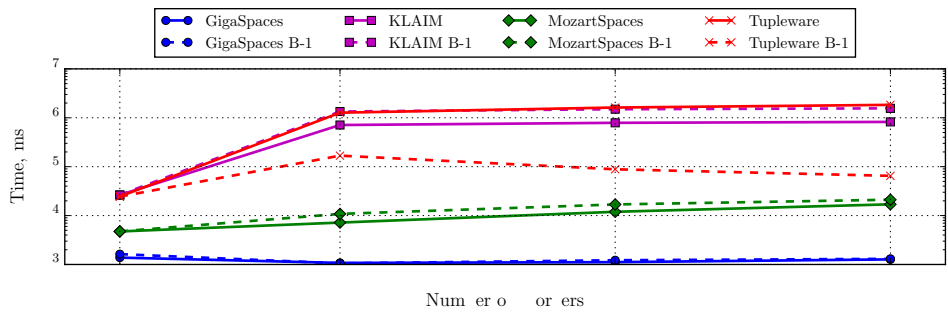

Figure 19. Matrix multiplication. Total time (matrix order 100)

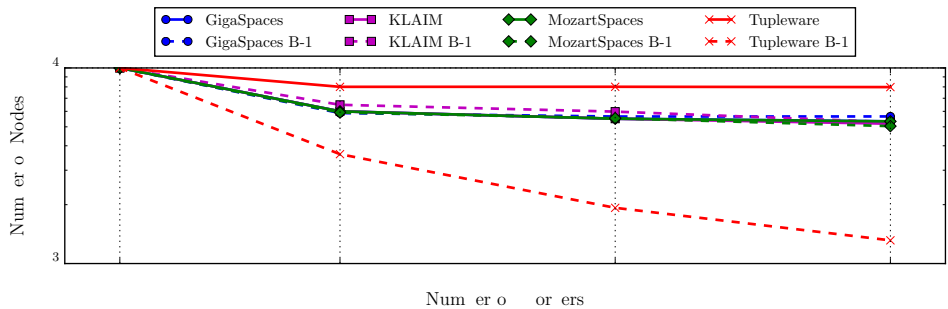

Figure 20. Matrix multiplication. Number of visited nodes (matrix order 100)

As shown in Figure 19, the runs with the uniform distribution outperform (i.e., they show a lower search time) the others, except the ones where TUPLEWARE is used. This is more evident in the case of KLAIM, where the difference in execution time is up to two times. To explain this behavior we looked at the time logs for one of the experiments (matrix order 50, workers 5) which consist of several files for each worker (e.g. Java Thread) and paid attention to the search time that mostly affects the execution time. The search time of each search operation performed during the execution of the case study for KLAIM and GIGASPACES is shown in Figure 23 and Figure 24 respectively (these two figures are presented not in a logarithmic scale). Every colored line represents one of five threads of workers and shows how the search time changes when the program executes. As we can see, although the search time of GIGASPACES is much less than the time of KLAIM, there is a specific regularity for both tuple space systems: the average value of search time for each thread significantly differs from each other. At the same time, the search operation is the most frequent one and it mostly defines the time of the execution. Therefore, a thread with the greatest average value of the search time determines the time of execution. The situation is similar with GigaSpaces and MozartSpaces, but less visible since the search time of Klaim is much greater (Figure 21).

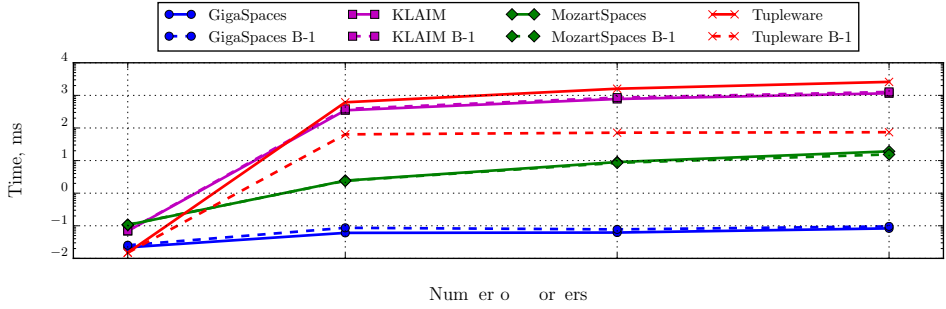

Figure 21. Matrix multiplication. Search time (matrix order 100)

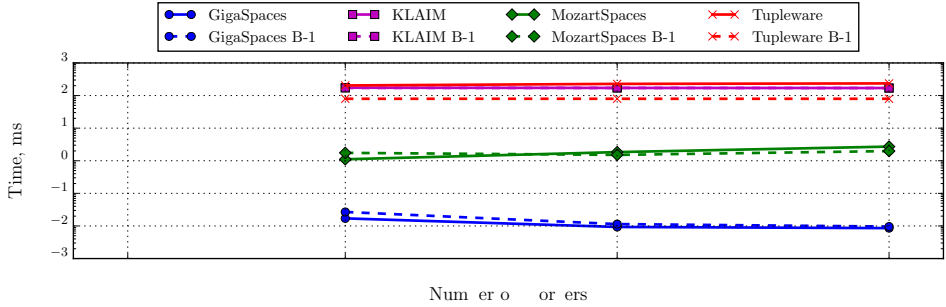

Figure 22. Matrix multiplication. Remote reading time (matrix order 100)

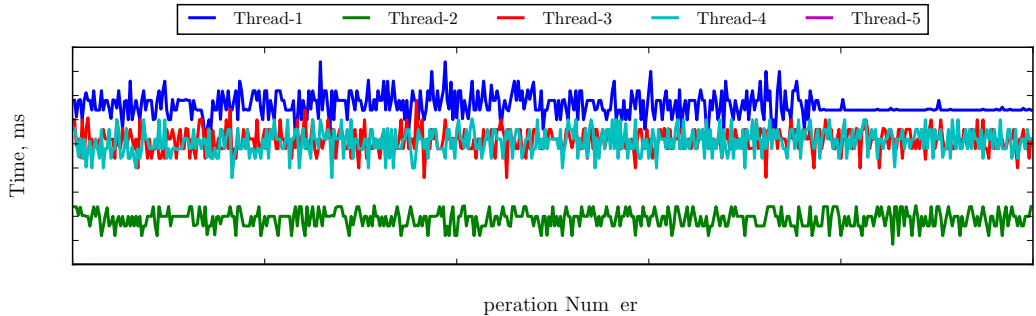

Figure 23. Matrix multiplication. Search time per thread (Klaim, matrix order 50, workers 5)

The results of this case study are generally consistent with the previous ones: remote operations of GigaSpaces and MozartSpaces are much faster and better fits to the application with frequent inter-process communication; Tupleware continues to have an advantage in the application with a specific pattern of communication. At the same time, we revealed that in some cases this feature of Tupleware had the side-effect that negatively affected its performance.

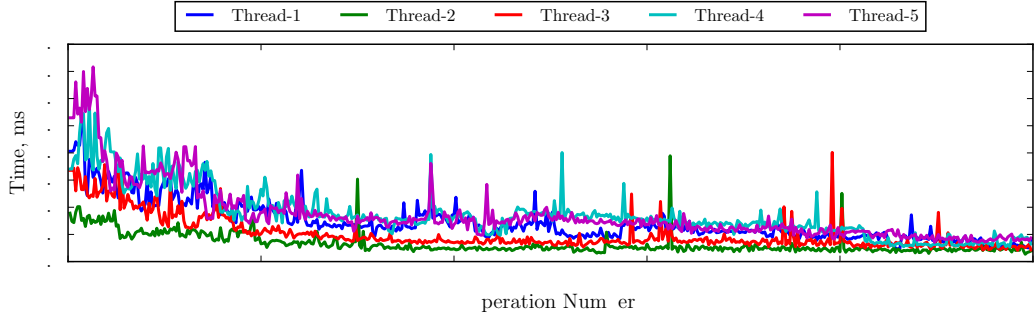

Figure 24. Matrix multiplication. Search time per thread (GigaSpaces, matrix order 50, workers 5)

**Modifying KLAIM.** The previous experiments provided us with the evidence that Klaim suffers from a high cost of communications when one process accesses the tuple space of another. At the same time, for instance, MozartSpaces does not have such a problem and, therefore, we have chosen to improve the part of the network communication in Klaim without touching others.

We then have substituted the part of Klaim responsible for sending and receiving tuples. It was based on Java IO, the package containing classes for the data transmission over the network. For the renewed part, we have opted to use Java NIO, non-blocking IO [29], which is a modern version of IO and in some cases allows for an efficient use of resources. Java NIO is beneficial when used to program applications dealing with many incoming connections. Moreover, for synchronization purposes, we used a more recent package (java.util.concurrent) instead of synchronization methods of the previous generation.

To evaluate the effectiveness of the modified Klaim we tested it with the *Matrix multiplication* case study since it depends on remote operations more than other case study and benefits of the modification are clearer. We just show the results for the case in which matrix B is uniformly distributed among the workers since the other case shows similar results. As shown in Figure 25 the remote writing time decreased significantly. The remote reading time of the modified Klaim is close to the one of MozartSpaces (Figure 26) and demonstrates similar behavior. In Figure 26 the remote reading time for the runs with one worker is not shown since in this case just the local tuple space of the worker is used. The remote reading operations mostly determine the total time and that is why graphics of modified Klaim and MozartSpaces in Figure 27 are similar.

Therefore, we have modified Klaim in order to decrease the time of interprocess communication and our changes of Klaim provide a significantly lower time of remote operations and lead to better performance of the tuple space system.

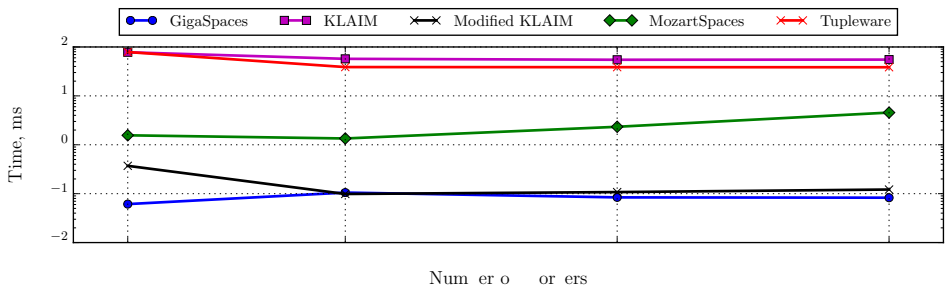

Figure 25. Matrix multiplication (with modified KLAIM).
Remote writing time (matrix order 100)

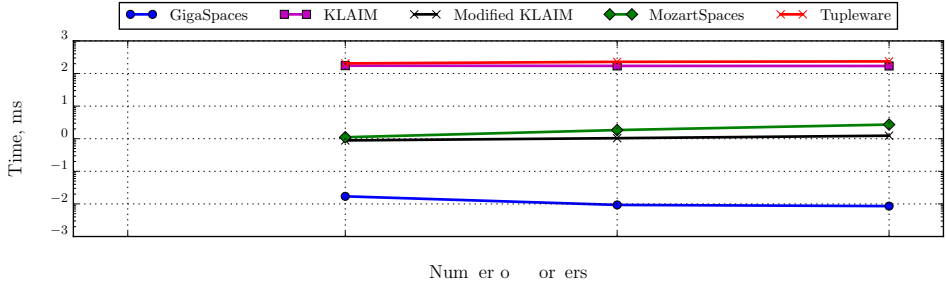

Figure 26. Matrix multiplication (with modified KLAIM).
Remote reading time (matrix order 100)

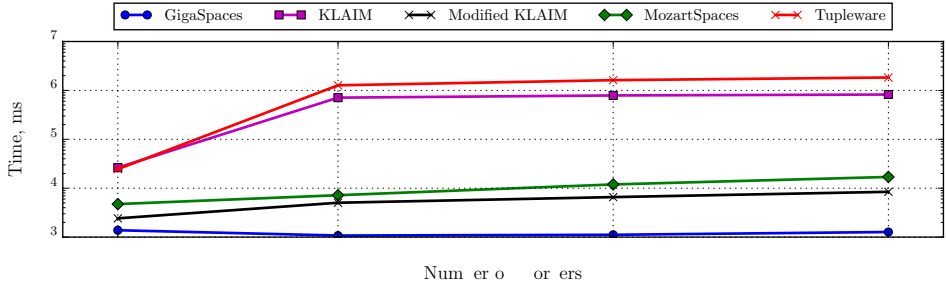

Figure 27. Matrix multiplication (with modified KLAIM). Total time (matrix order 100)

**Experiments with several host machines.** The results of the previous experiments which were conducted using only one host machine provide us evidence that GigaSpaces has a more efficient implementation of communication and that is very beneficial when many operations on remote tuple spaces are used. Since we do not have access to its source code we conjecture that GigaSpaces uses an efficient inter-process communicating mechanism and do not resort to socket communications as the other implementations. To check whether GigaSpaces continues to be so efficient when the networking has to be used, we have created a network of several identical host machines using Oracle VM VirtualBox[§]. The configuration of the hosts differs from the one of the previous experiments: 1 CPU, 1 GB RAM and installed Linux Mint 17.3 Xfce. The master and each worker process were launched in their own hosts. Each set of experiments was conducted 10 times and after the results of the execution were collected and analyzed.

We conducted experiments for two case studies: *Sorting* and *Matrix multiplication* (an implementation with the uniform distribution). We just focused on remote reading time, since it is the most frequently used operations in these two case studies. In all the tests for the networked version, the remote reading time exceeds significantly the time measured in one host version showing that for the one host case study implementation GigaSpaces does not use network protocols. In addition, by comparing two case studies we can notice the following. First, we compare the ratio of the remote reading time of networked version to one of one host version and for two different case studies the ratio was completely different: for *Sorting* it is around 20 (Table II), for *Matrix multiplication* it is around 100 (Table III). This discrepancy is related to the difference in type and size of transmitted data. Second, the use of several separate hosts affects the total time differently: for instance, considering *Sorting*, that leads to the acceleration of the execution (Table II). The reasons of that are not clear and related to the internal implementation of GigaSpaces.

---

[§]Oracle VM VirtualBox is a free and open-source hypervisor for x86 computers from Oracle Corporation (https://www.virtualbox.org/).

|  | **Remote reading time** | **Search time** | **Total time** |
|---|---|---|---|
| **1M elements** | | | |
| Single host version | 0.1247 | 1.1324 | 1259.0 |
| Networked version | 2.7620 | 9.7291 | 2706.0 |
| **10M elements** | | | |
| Single host version | 0.1939 | 12.9030 | 84461.0 |
| Networked version | 3.8184 | 91.9117 | 50728.0 |

Table II. Sorting. Comparison single host version and networked version (workers 5)

|  | **Remote reading time** | **Search time** | **Total time** |
|---|---|---|---|
| Single host version | 0.0171 | 0.0606 | 1076.0 |
| Networked version | 1.6240 | 4.1299 | 17354.0 |

Table III. Matrix multiplication. Comparison single host version and networked version (matrix order 5, workers 5)

## 4. CONCLUSIONS

Distributed computing is getting increasingly pervasive, with demands from various applications domains and highly diverse underlying architectures from the multitude of tiny things to the very large cloud-based systems. Tuple spaces certainly offer valuable tools and methodologies to help develop scalable distributed applications/systems. This paper has first surveyed and evaluated a number of tuple space systems, then it has analyzed more closely four different systems. We considered GigaSpaces, because it is one of the few currently used commercial products, Klaim, because it guarantees code mobility and flexible manipulation of tuple spaces, MozartSpaces as the most recent implementation that satisfies the main criteria we do consider essential for tuple based programming, and Tupleware, because it is the one that turned out to be the best in our initial evaluation. We have then compared the four system by evaluating their performances over four case studies: one testing performance of local tuple space, a communication-intensive one, a computational-intensive one, and one demanding a specific communication pattern.

Our work follows the lines of [34] but we have chosen more recent implementations and conducted more extensive experiments. On purpose, we ignored implementations of systems that have been directly inspired by those considered in the paper. Thus, we did not consider

jRESP[¶] a Java runtime environment, that provides a tool for developing autonomic and adaptive systems according to the SCEL approach [16, 17].

After analyzing the outcome of the experiments, it became clear what are the critical aspects of a tuple space system that deserve specific attention to obtain efficient implementations. The critical choices are concerned with inter-process and inter-machine communication and local tuple space management.

The first aspect is related to data transmission and is influenced by the choice of algorithms that reduce communication. For instance, the commercial system GIGASPACES differs from the other systems that we considered for the technique used for data exchange, exploiting memory based inter-process communication, that guarantees a considerably smaller access time to data. Therefore, the use of this mechanism on a single machine does increase efficiency. However, when working with networked machines, it is not possible to use the same mechanism and we need to resort to other approaches (e.g. the TUPLEWARE one) to reduce inter-machine communication and to have more effective communications. To compare GIGASPACES with the other tuple space systems under similar conditions and thus to check whether it remains efficient also in the case of distributed computing, we have carried out experiments using a network where workers and masters processes are hosted on different machines. The results of these experiments show that, although the remote operations are much slower, the overall performance of GIGASPACES remains high.

The second aspect, concerned with the implementation of local tuple spaces, is heavily influenced by the data structure chosen to represent tuples, the corresponding data matching algorithms and by the lock mechanisms used to prevent conflicts when accessing the tuple space. In our experiments the performance of different operations on tuple spaces varies considerably; for example, KLAIM provides fast writing and slow reading, whereas TUPLEWARE exhibits high writing time and fast reading time. The performances of a tuple space system would depend also on the chosen system architectures which determine the kind of interaction between their components. Indeed, it is evident that all the issues should be tackled together because they are closely interdependent. Another interesting experiment would be to use one of the classical database systems that offer fast I/O operations to model tuple spaces and their operations, in order to assess their performances over our case studies. The same experiment can be carried out by considering modern no-sql databases, such as Redis[∥] or MongoDB[**].

Together with the experiments, we have started modifying the implementation of one of the considered systems, namely KLAIM. We have focused on the part which evidently damages its overall performance, i.e., the one concerned with data transmission over the network. Our experiments have shown an interesting outcome. The time of remote writing and reading becomes comparable to that of MOZARTSPACES and is significantly shorter than that required by the previous implementation of KLAIM. Indeed, the modified version allows much faster execution of the tasks where many inter-process communications are considered.

We plan to use the results of this work as the basis for designing an efficient tuple space system which offers programmers the possibility of selecting (e.g. via a dashboard) the desired features of the tuple space according to the specific application. In this way, one could envisage a distributed middleware with different tuple spaces implementations each of them targeted to specific classes of systems and devised to guarantee the most efficient execution. The set of configuration options will be a key factor of the work. One of such options, that we consider important for improving performances is data replication. In this respect, we plan to exploit the results of RepliKlaim [2] which enriched KLAIM with primitives for replica-aware coordination. Indeed, we will use the current implementation of KLAIM as a starting point for the re-engineering of tuple space middleware.

---

[¶]Documentation and the source code for jRESP are available at http://www.jresp.sourceforge.net/
[∥]https://redis.io/
[**]https://www.mongodb.com/

ACKNOWLEDGEMENT

The authors would like to thank Lorenzo Bettini and Michele Loreti for fruitful discussions on tuple spaces implementations, and Alberto Lluch Lafuente for suggestions and feedback on a preliminary version of this document.

REFERENCES

1. G. R. Andrews. *Concurrent Programming: Principles and Practice*. Benjamin-Cummings Publishing Co., Inc., USA, 1991.
2. M. Andric, R. De Nicola, and A. Lluch-Lafuente. Replica-based high-performance tuple space computing. In T. Holvoet and M. Viroli, editors, *Coordination Models and Languages - COORDINATION 2015*, volume 9037 of *Lecture Notes in Computer Science*, pages 3–18. Springer, 2015.
3. A. Atkinson. *Tupleware: A Distributed Tuple Space for the Development and Execution of Array-based Applications in a Cluster Computing Environment*. University of Tasmania School of Computing and Information Systems thesis. University of Tasmania, 2010.
4. R. R. Atkinson and C. Hewitt. Parallelism and synchronization in actor systems. In R. M. Graham, M. A. Harrison, and R. Sethi, editors, *Conference Record of the Fourth ACM Symposium on Principles of Programming Languages, USA*, pages 267–280. ACM, 1977.
5. B. Barker. Message Passing Interface (MPI). In *Workshop: High Performance Computing on Stampede*, 2015.
6. L. Bettini, R. De Nicola, and M. Loreti. Implementing mobile and distributed applications in X-Klaim. *Scalable Computing: Practice and Experience*, 7(4), 2006.
7. A. Birrell and B. J. Nelson. Implementing remote procedure calls. *ACM Trans. Comput. Syst.*, 2(1):39–59, 1984.
8. V. Buravlev, R. De Nicola, and C. A. Mezzina. Tuple spaces implementations and their efficiency. In A. Lluch-Lafuente and J. Proença, editors, *Coordination Models and Languages - COORDINATION 2016*, volume 9686 of *Lecture Notes in Computer Science*, pages 51–66. Springer, 2016.
9. S. Capizzi. *A Tuple Space Implementation for Large-Scale Infrastructures*. Department of Computer Science Univ. Bologna thesis. University of Bologna, 2008.
10. N. Carriero and D. Gelernter. *How to write parallel programs - a first course*. MIT Press, 1990.
11. S. Cass. The 2015 top ten programming languages. *IEEE Spectrum*, 2015.
12. M. Ceriotti, A. L. Murphy, and G. P. Picco. Data sharing vs. message passing: Synergy or incompatibility? an implementation-driven case study. *Proceedings of the 2008 ACM symposium on Applied computing*, pages 100–107, 2008.
13. S. Craß, T. Dönz, G. Joskowicz, eva Kühn, and A. Marek. Securing a space-based service architecture with coordination-driven access control. *JoWUA*, 4(1):76–97, 2013.
14. S. Craß, eva Kühn, and G. Salzer. Algebraic foundation of a data model for an extensible space-based collaboration protocol. In *International Database Engineering and Applications Symposium (IDEAS 2009)*, pages 301–306, 2009.
15. R. De Nicola, G. L. Ferrari, and R. Pugliese. KLAIM: A kernel language for agents interaction and mobility. *IEEE Trans. Software Eng.*, 24(5):315–330, 1998.
16. R. De Nicola, D. Latella, A. Lluch-Lafuente, M. Loreti, A. Margheri, M. Massink, A. Morichetta, R. Pugliese, F. Tiezzi, and A. Vandin. The SCEL language: Design, implementation, verification. In M. Wirsing, M. Hlzl, N. Koch, and P. Mayer, editors, *Software Engineering for Collective Autonomic Systems - The ASCENS Approach*, volume 8998 of *Lecture Notes in Computer Science*, pages 3–71. Springer, 2015.
17. R. De Nicola, M. Loreti, R. Pugliese, and F. Tiezzi. A formal approach to autonomic systems programming: The SCEL language. *TAAS*, 9(2):7:1–7:29, 2014.
18. P. T. Eugster, P. Felber, R. Guerraoui, and A. Kermarrec. The many faces of publish/subscribe. *ACM Comput. Surv.*, 35(2):114–131, 2003.
19. D. Gelernter. Generative communication in Linda. *ACM Trans. Program. Lang. Syst.*, 7(1):80–112, 1985.
20. D. Gelernter and A. J. Bernstein. Distributed communication via global buffer. In *ACM SIGACT-SIGOPS Symposium on Principles of Distributed Computing, 1982*, pages 10–18, 1982.
21. GigaSpaces. Concepts - XAP 9.0 Documentation - GigaSpaces Documentation Wiki. `http://wiki.gigaspaces.com/wiki/display/XAP9/Concepts`. [Online; accessed 15-September-2016].
22. Y. Jiang, Z. Jia, G. Xue, and J. You. Dtuples: A distributed hash table based tuple space service for distributed coordination. *Grid and Cooperative Computing, 2006. GCC 2006. Fifth International Conference*, pages 101–106, 2006.
23. M. A. Leal, N. Rodriguez, and R. Ierusalimschy. Luats a reactive event-driven tuple space. *Journal of Universal Computer Science*, 9(8):730–744, 2003.
24. T. Lehman, S. McLaughry, and P. Wyckoff. Tspaces: The next wave. In *HICSS 1999: Proceedings of the Thirty-second Annual Hawaii International Conference on System Sciences*, volume 8. Springer Berlin Heidelberg, 1999.
25. T. Mattson, B. Sanders, and B. Massingill. *Patterns for Parallel Programming*. The software patterns series. Addison-Wesley, 2005.
26. S. Microsystems. JS - JavaSpaces Service Specification. `https://river.apache.org/doc/specs/html/js-spec.html`. [Online; accessed 15-September-2016].
27. R. Mordinyi, eva Kühn, and A. Schatten. Space-based architectures as abstraction layer for distributed business applications. In *CISIS 2010, The Fourth International Conference on Complex, Intelligent and*

*Software Intensive Systems*, pages 47–53. IEEE Computer Society, 2010.
28. B. Nitzberg and V. M. Lo. Distributed shared memory: A survey of issues and algorithms. *IEEE Computer*, 24(8):52–60, 1991.
29. Oracle. Package java.nio. `https://docs.oracle.com/javase/7/docs/api/java/nio/package-summary.html`. [Online; accessed 15-September-2016].
30. E. Pitt and K. McNiff. *java.rmi: The Remote Method Invocation Guide*. Addison-Wesley Longman Publishing Co., Inc., USA, 2001.
31. M. J. Quinn. *Parallel Programming in C with MPI and OpenMP*. McGraw-Hill Higher Education, 2004.
32. TIOBE index web site. `http://www.tiobe.com/tiobe-index/`. [Online; accessed 15-September-2016].
33. R. Van Der Goot. *High Performance Linda Using a Class Library*. PhD thesis. Erasmus Universiteit Rotterdam, 2001.
34. G. Wells, A. Chalmers, and P. G. Clayton. Linda implementations in java for concurrent systems. *Concurrency - Practice and Experience*, 16(10):1005–1022, 2004.

# Appendices

## A. PASSWORD SEARCH LISTINGS

Listing 1 shows that the worker process also starts its local tuple space and checks the connection to the master and other workers (lines 27-31). Then the process loads predefined data into its local space (line 38) and taking tasks from the master begins to search for necessary tuples in its local and remote tuple spaces (lines 58-78). At the end, when all tasks are accomplished the worker saves the data of profiling ((line 76)).

Listing 1: Password search. Listing of the worker process

```java
public class DistributedSearchWorker<T extends ITupleSpace> {

  // class fields
  private Object localTSAddress;
  Object masterTSAddress;
  ArrayList<Object> otherWorkerTSName;
  ArrayList<ITupleSpace> workerTSs;
  Integer workerID;
  int matrixSize;
  Integer numberOfWorkers;
  Class tupleSpaceClass;

  public DistributedSearchWorker(Object localTSAddress, Integer workerID,
      Object masterTSAddress,
    ArrayList<Object> otherWorkerTSName, int matrixSize, int numberOfWorkers,
        Class tupleSpaceClass) {
    //initialising fields
  }

  /***
  * description of the worker process
  * @throws NoSuchAlgorithmException
  * @throws IOException
  * @throws InterruptedException
  */
  public void passwordSearchWorker() throws NoSuchAlgorithmException,
      IOException, InterruptedException
  {
    // initialize a local tuple space
    T localTS = getInstanceOfT(tupleSpaceClass);
    localTS.startTupleSpace(localTSAddress, numberOfWorkers, true);

    // connect to the master and to the other workers
    T masterTS = initializeMasterAndWorkers();

    TupleOperations.writeTuple(masterTS, localTS,
        localTS.formTuple("SearchTuple",
      new Object[]{"search", "worker", "worker_ready"},
      DistributedSearchMaster.searchTupleTemplate), false, true);

    // load data to the local tuple space
    loadDataTable(localTS);

    // notify master that worker is ready to receive tasks
    TupleOperations.writeTuple(masterTS, localTS,
        localTS.formTuple("SearchTuple",
      new Object[]{"search", "worker", "data_loaded"},
      DistributedSearchMaster.searchTupleTemplate), false, true);

    // get test key of the execution
    Object tupleWithTestKeyObject = TupleOperations.readTuple(masterTS,
        localTS,
      localTS.formTuple("SearchTuple", new Object[]{"search", "master_key",
          null},
      DistributedSearchMaster.searchTupleTemplate), false, true);
    Object[] tupleWithTestKey = localTS.tupleToObjectArray("SearchTuple",
        tupleWithTestKeyObject);
    DProfiler.testKey = (String)tupleWithTestKey[2];
```

```java
51
52          // get a task from the master
53          Object[] searchTaskTuple = searchNextTask(localTS, masterTS);
54
55          String hashedValue = (String)searchTaskTuple[1];
56          String status = (String)searchTaskTuple[2];
57
58          while(!GigaspacesDistPassSearchMaster.completeStatus.equals(status))
59          {
60            // search for the tuple with given hashed value
61            Object foundTupleObject = searchLoop(localTS, workerTSs, hashedValue);
62            Object[] foundTuple = localTS.tupleToObjectArray("SearchTuple",
                  foundTupleObject);
63
64            // send found password to the master
65            TupleOperations.writeTuple(masterTS, localTS,
                  localTS.formTuple("SearchTuple",
66              new Object[]{"foundValue", (String)foundTuple[1],
                    (String)foundTuple[2]},
67              DistributedSearchMaster.searchTupleTemplate), false, true);
68
69            // search and retrieve the next task
70            searchTaskTuple = searchNextTask(localTS, masterTS);
71            hashedValue = (String)searchTaskTuple[1];
72            status = (String)searchTaskTuple[2];
73          }
74
75        // write data of the test
76        TupleLogger.writeAllToFile(DProfiler.testKey);
77        localTS.stopTupleSpace();
78      }
79
80        /***
81         * initialize connections with master and workers
82         * @return
83         */
84      private T initializeMasterAndWorkers() {
85        // trying to connect to server tuple space 5 times
86            T masterTS = getInstanceOfT(tupleSpaceClass);
87            masterTS.startTupleSpace(masterTSAddress, numberOfWorkers, false);
88
89            // create connection to other workers
90            workerTSs = new ArrayList<>();
91            Collections.shuffle(otherWorkerTSName);
92            for(int i = 0; i < otherWorkerTSName.size(); i++)
93            {
94              T workerTS = getInstanceOfT(tupleSpaceClass);
95              workerTS.startTupleSpace(otherWorkerTSName.get(i), numberOfWorkers,
                    false);
96              workerTSs.add(workerTS);
97            }
98        return masterTS;
99      }
100
101     /***
102      * load data with passwords from the file
103      * @param localTS local tuple space
104      * @throws IOException
105      */
106     private void loadDataTable(T localTS) throws IOException {
107       String[] dataArray = GigaspacesDistSearchTest.getStringArray("hashSet" +
              workerID + ".dat");
108         for(int i =0; i< dataArray.length; i++)
109         {
110           String[] elements = dataArray[i].split(",");
111           // write data to the local tuple space
112             TupleOperations.writeTuple(localTS, localTS,
                  localTS.formTuple("SearchTuple", new Object[]{"hashSet",
                    elements[0], elements[1]},
                    DistributedSearchMaster.searchTupleTemplate), true, true);
113         }
```

```java
114      }
115
116      /***
117       * get next task from the master
118       * @param localTS   local tuple space
119       * @param masterTS master tuple space
120       * @return
121       */
122      private Object[] searchNextTask(T localTS, T masterTS) {
123         Object searchTaskTupleTemplate = localTS.formTuple("SearchTuple", new
                 Object[]{"search_task", null, null},
                 DistributedSearchMaster.searchTupleTemplate);
124         Object searchTaskTupleObject = TupleOperations.takeTuple(masterTS,
                 localTS, searchTaskTupleTemplate, false, true);
125         Object[] searchTaskTuple = localTS.tupleToObjectArray("SearchTuple",
                 searchTaskTupleObject);
126         return searchTaskTuple;
127      }
128
129      /***
130       * continuously search a password
131         * @param localTS      local tuple space
132         * @param workerTSs    tuple spaces of other workers
133         * @param hashedValue hashed value for the search
134       * @return
135       * @throws InterruptedException
136       */
137        static Object searchLoop(ITupleSpace localTS, ArrayList<ITupleSpace>
                 workerTSs, String hashedValue) throws InterruptedException
138       {
139          boolean firstTime = true;
140       TupleLogger.begin("read::l-r");
141          Object result = null;
142          while(true)
143          {
144             result = search(localTS, workerTSs, hashedValue, firstTime);
145             firstTime = false;
146             if(result != null)
147             {
148                TupleLogger.end("read::l-r");
149                return result;
150             }
151             try {
152             Thread.sleep(1);
153          } catch (InterruptedException e) {
154             e.printStackTrace();
155          }
156          }
157       }
158
159      /***
160       * search in the distribution of tuple spaces
161       * @param localTS      local tuple space
162       * @param workerTSs    tuple spaces of other workers
163       * @param hashedValue hashed value for the search
164       * @param firstTime
165       * @return
166       * @throws InterruptedException
167       */
168       static Object search(ITupleSpace localTS, ArrayList<ITupleSpace>
                 workerTSs, String hashedValue, boolean firstTime) throws
                 InterruptedException
169      {
170         Object template = localTS.formTuple("SearchTuple", new
                 Object[]{"hashSet", hashedValue, null},
                 DistributedSearchMaster.searchTupleTemplate);
171         if(firstTime)
172            TupleLogger.incCounter("nodeVisited");
173
174         // read from local space
175         Object resultTuple = null;
```

```
176        if (localTS instanceof TuplewareProxy)
177          resultTuple = TupleOperations.readIfExistTuple(localTS, localTS,
                 template, true, false);
178        else
179          resultTuple = TupleOperations.readIfExistTuple(localTS, localTS,
                 template, true, true);
180
181        if (resultTuple != null)
182          return resultTuple;
183        else
184        {
185          // search in tuple spaces of the other workers
186            for(int i = 0; i< workerTSs.size(); i++)
187            {
188                if(firstTime)
189                   TupleLogger.incCounter("nodeVisited");
190              // read from remote space
191                    resultTuple = TupleOperations.readIfExistTuple(workerTSs.get(i),
                       localTS, template, false, true);
192                if(resultTuple != null)
193                   return resultTuple;
194            }
195        }
196        return null;
197        }
198
199     /***
200      * create an object using class name
201      * @param aClass class name
202      * @return
203      */
204      public T getInstanceOfT(Class<T> aClass)
205      {
206          try {
207          return aClass.newInstance();
208        } catch (InstantiationException | IllegalAccessException e) {
209          e.printStackTrace();
210        }
211          return null;
212        }
213 }
```

Listing 2: Password search. Listing of the master process

```
1  public class DistributedSearchMaster<T extends ITupleSpace> {
2
3    // class fields
4    private Object masterTSAddress;
5    private int numberOfWorkers;
6    private int numberOfElements;
7    private Class tupleSpaceClass;
8
9    // tuple type
10   public static Object[] searchTupleTemplate =
11     new Object[] {String.class, String.class, String.class};
12
13   public DistributedSearchMaster(Object masterTSAddress, int numberOfElements,
14     int numberOfWorkers, Class tupleSpaceClass) {
15     //initialising fields
16   }
17
18     /***
19      * description of the master process
20      * @throws NoSuchAlgorithmException
21      * @throws InterruptedException
22      * @throws FileNotFoundException
23      * @throws IOException
24      */
25    public void passwordSearchMaster()  {
26
27        // initialize a local tuple space
```

```java
28         T masterTS = getInstanceOfT(tupleSpaceClass);
29         masterTS.startTupleSpace(masterTSAddress, numberOfWorkers, true);
30
31         // wait when all workers will be available
32         waitForAllWorkers(masterTS);
33         TupleLogger.begin("Master::TotalRuntime");
34
35         // wait when all workers will load tables with data
36         waitForDataLoad(masterTS);
37
38         // spread the current test key
39         TupleOperations.writeTuple(masterTS, masterTS,
               masterTS.formTuple("SearchTuple",
40           new Object[]{"search", "master_key", DProfiler.testKey},
41           searchTupleTemplate), true, false);
42
43         // create tasks and write it into local tuple space of the master
44         int numberOfTasks = 100;
45         taskCreation(masterTS, numberOfTasks);
46
47         // wait and get accomplished tasks
48         ArrayList<Object[]> foundTuples = getAccomplishedTasks(masterTS,
               numberOfTasks);
49
50         // send to worker - "finish its work"
51         for(int i =0; i< numberOfWorkers; i++) {
52         TupleOperations.writeTuple(masterTS, masterTS,
53           masterTS.formTuple("SearchTuple", new Object[]{"search_task", "",
               completeStatus},
54           searchTupleTemplate), true, false);
55         }
56         TupleLogger.end("Master::TotalRuntime");
57
58         // write data of the test
59         DProfiler.writeTestKeyToFile(DProfiler.testKey);
60         TupleLogger.writeAllToFile(DProfiler.testKey);
61
62      // wait until all workers will end
63         Thread.sleep(10000);
64         masterTS.stopTupleSpace();
65      }
66
67   /***
68    * wait when all workers are available
69    * @param masterTS the master tuple space
70    * @throws InterruptedException
71    */
72   private void waitForAllWorkers(T masterTS) throws InterruptedException {
73      int workerCounter = 0;
74      while (true) {
75         Thread.sleep(10);
76         TupleOperations.takeTuple(masterTS, masterTS,
               masterTS.formTuple("SearchTuple", new Object[]{"search", "worker",
               "worker_ready"}, searchTupleTemplate), true, false);
77         workerCounter++;
78         if(workerCounter == numberOfWorkers)
79         break;
80      }
81      System.out.println("Master: all worker loaded data");
82   }
83
84   /***
85    * wait when all workers load tables with data
86    * @param masterTS the master tuple space
87    * @throws InterruptedException
88    */
89   private void waitForDataLoad(T masterTS) throws InterruptedException {
90      int workerCounter;
91      workerCounter = 0;
92      while (true) {
93         Thread.sleep(10);
```

```java
94          TupleOperations.takeTuple(masterTS, masterTS,
                masterTS.formTuple("SearchTuple", new Object[]{"search", "worker",
                "data_loaded"}, searchTupleTemplate), true, false);
95          workerCounter++;
96          if(workerCounter == numberOfWorkers)
97            break;
98        }
99     }
100
101    /***
102     * create tasks and write it into local tuple space of the master
103     * @param masterTS       the master tuple space
104     * @param numberOfTasks  the number of tasks
105     * @throws NoSuchAlgorithmException
106     */
107    private void taskCreation(T masterTS, int numberOfTasks) throws
            NoSuchAlgorithmException {
108      Random r = new Random();
109      MessageDigest mdEnc = MessageDigest.getInstance("MD5");
110      String[] tasks = new String[numberOfTasks];
111      for(int i = 0; i < tasks.length; i++) {
112        int nextNumber = r.nextInt(numberOfElements);
113        tasks[i] = GigaspacesDistSearchTest.integerToHashString(mdEnc,
              nextNumber);
114      }
115
116      // put all tasks into tupleSpace
117      for(int i = 0; i < tasks.length; i++) {
118        TupleOperations.writeTuple(masterTS, masterTS,
              masterTS.formTuple("SearchTuple", new Object[]{"search_task",
              tasks[i], "not_processed"}, searchTupleTemplate), true, false);
119      }
120      System.out.println("Master: tasks were created");
121    }
122
123    /***
124     * get accomplished tasks
125     * @param masterTS       the master tuple space
126     * @param numberOfTasks  the number of tasks
127     * @return
128     */
129    private ArrayList<Object[]> getAccomplishedTasks(T masterTS, int
            numberOfTasks) {
130      ArrayList<Object[]> foundTuples = new ArrayList<Object[]>();
131      while (foundTuples.size() != numberOfTasks) {
132        Object foundTupleTemplate = masterTS.formTuple("SearchTuple", new
              Object[]{"foundValue", null, null}, searchTupleTemplate);
133        Object foundTupleObject = TupleOperations.takeTuple(masterTS, masterTS,
              foundTupleTemplate, true, false);
134        Object[] foundTuple = masterTS.tupleToObjectArray("SearchTuple",
              foundTupleObject);
135        foundTuples.add(foundTuple);
136      }
137      return foundTuples;
138    }
139
140    /***
141     * create an object using class name
142     * @param aClass class name
143     * @return
144     */
145    public T getInstanceOfT(Class<T> aClass) {
146      try {
147        return aClass.newInstance();
148      } catch (InstantiationException | IllegalAccessException e) {
149        e.printStackTrace();
150      }
151      return null;
152    }
153  }
```

Listing 3: Password search. Listing of the method for the starting of master and worker threads

```java
private static void passwordSearchGigaSpaces(int numberOfElements, int
    numberOfWorkers) {

  // define addresses of the master and workers
  String basicAddress = "/./";
  String masterAddress = basicAddress + "master";
  ArrayList<String> allWorkers = new ArrayList<String>();
  for(int k = 0; k < numberOfWorkers; k++)
    allWorkers.add(basicAddress + "worker" + k);

  // start worker threads
  for(int i=0; i < numberOfWorkers; i++)
  {
    ArrayList<Object> otherWorkerTSName = new ArrayList<Object>();
    for(int k = 0; k < numberOfWorkers; k++)
    {
      if(i != k)
        otherWorkerTSName.add(allWorkers.get(k));
    }
    DistributedSearchWorkerThread<GigaSpaceProxy> wThread =
      new DistributedSearchWorkerThread<GigaSpaceProxy>(allWorkers.get(i), i,
          masterAddress,
        otherWorkerTSName, numberOfElements, numberOfWorkers,
            GigaSpaceProxy.class);
    wThread.start();
  }

  // start the master thread
  DistributedSearchMasterThread<GigaSpaceProxy> mThread =
    new DistributedSearchMasterThread<GigaSpaceProxy>(masterAddress,
        numberOfElements,
      numberOfWorkers, GigaSpaceProxy.class);
  mThread.start();
}
```